\documentclass[twocolumn]{emulateapj}
\usepackage{amsmath}
\usepackage{hyperref}
\usepackage{graphicx}                                                                                                                                                                                                                                                                                                                                                                                                                                                                                                                                                                                                                                                                                                                                                                                                                                                                                                                                                                                                                                                      
\def\galaxiespprox{\mathrel{\vcenter{\offinterlineskip \hbox{$>$}
    \kern 0.3ex \hbox{$\sim$}}}}
\def\lapprox{\mathrel{\vcenter{\offinterlineskip \hbox{$<$}
    \kern 0.3ex \hbox{$\sim$}}}}
\newcommand{\beq}{\begin{equation}} 
\newcommand{\eeq}{\end{equation}}

\def\Sig15{\Sigma_{1.5}}

\def\Msun{\hbox{$\thinspace \rm M_{\odot}$}}

\def\delm12{M_{12}}
\def\sigm1{\sigma(m_{1})}

\begin{document}

\shortauthors{Ostriker et al.}
\shorttitle{Mind The Gap}

\title{MIND THE GAP: Is The Too Big To Fail Problem Resolved?}
\author{Jeremiah P. Ostriker\altaffilmark{1,2},
             Ena Choi\altaffilmark{1}, 
              Anthony Chow\altaffilmark{1}, 
              Kundan Guha\altaffilmark{1}}
              
\affil{ $^1$Department of Astronomy, Columbia University, New York, NY 10027, USA \\
        $^2$ Department of Astrophysical Sciences, Princeton University, Princeton, 
                 NJ 08544, USA\\
  \texttt{jpo@astro.columbia.edu} \\}

\begin{abstract}
The faintness of satellite systems in galaxy groups has contributed to the 
widely discussed ``missing satellite'' and ``too big to fail'' issues. Using
techniques based on \citet{1977ApJ...212..311T}, we show that there
is no problem with the luminosity function computed from modern 
codes per se, but that the gap between first and second brightest 
systems is too big {\it given} the luminosity function, that the same large 
gap is found in modern, large scale baryonic $\Lambda$CDM 
simulations such as EAGLE and IllustrisTNG, is even greater in dark 
matter only simulations, and finally, that this is most likely due to 
gravitationally induced merging caused by classical dynamical friction. 
Quantitatively the gap is larger in the computed simulations than in 
the randomized ones by $1.79 \pm 1.04$, $1.51 \pm 0.93$,  
$3.43 \pm 1.44$ and $3.33 \pm 1.35$ magnitudes in the EAGLE, 
IllustrisTNG, and dark matter only simulations of EAGLE and IllustrisTNG 
respectively. Furthermore the anomalous gaps in the simulated systems 
are even larger than in the real data by over half a magnitude and are 
still larger in the dark matter only simulations. Briefly stated, 
$\Lambda$CDM does not have a problem with an absence of ``too big 
to fail'' galaxies. Statistically significant large gaps between first and 
second brightest galaxies are to be expected.
\end{abstract}

\section{Introduction}\label{intro}
There are two frequently discussed ``problems'' found in galaxy statistics
which are sometimes considered arguments against the standard 
$\Lambda$CDM model of cosmology. Both are related to the apparent
under-abundance of faint, low mass galaxies in local groups. One, ``the missing
satellite problem'' \citep{1993MNRAS.264..201K,1999ApJ...522...82K,
1999ApJ...524L..19M} notes that the CDM subhalo 
stellar mass function is steeper than the observed satellite mass function. The 
second, ``the too big to fail problem (TBTF)'' \citep[e.g.][]{2011MNRAS.415L..40B,
2012MNRAS.422.1203B,2014MNRAS.444..222G}
notes that {\it given} the observed stellar mass function, there should be many
intermediate mass systems in the local group and other nearby systems
that are missing.  The original paper, which introduced the TBTF issue, focused
primarily on the gap between the third and forth brightest galaxies in the local
group but most subsequent work has focussed on how much brighter the first 
brightest galaxy is than its companions. A recent paper entitled ``A Lonely Giant'' 
\citep{2018ApJ...863..152S} focusses on the under-abundance of
moderate mass satellite galaxies in the nearby M94 system. 

Both the nature and the significance of the two ``problems'' are often
confused. The ``missing satellite problem'' is an expression of our
surprise that the mass function for galaxies at the faint end is 
significantly less steep than the mass function expected for dark matter halos
-- if the CDM model is correct. It also implies that either CDM produces
too many low mass halos/subhalos or galaxies form in these halos
with lower and lower efficiency as the halo mass declines. Prevailing 
expert views at present seem to prefer the second explanation, and current
high quality simulations based on the CDM paradigm do, in fact, 
produce the correct luminosity function \citep[e.g.][]{2015MNRAS.446..521S,
2018MNRAS.473.4077P,2019MNRAS.tmp..904D}.

However, using the {\it observed} luminosity function (or the one computed 
with appropriate baryonic physics) it is hard to understand the
faintness of satellite systems in well observed groups and clusters in 
comparison to the first brightest system; that is the ``too big to fail problem'': 
there are relatively bright galaxies that are expected to be present which are 
among the missing.

What are missing here are moderate
mass galaxies roughly one or two magnitudes fainter than the brightest
central galaxy. The problem shows up
to observers as a large gap between the brightness of first and second
brightest galaxies in groups and clusters. These anomalously large
gaps were noticed as far back as \citet{1973ApJ...183..743S} and 
\citet{1978ApJ...223..765D}.

However, there is a 
brilliant paper by \citet{1977ApJ...212..311T} which sheds a blazing
light on the issues and makes clear that there must be interactions
amongst the group galaxies to be considered and that the gap between
first and second brightest galaxies is too big {\it given} the luminosity function.
The problems are not with the luminosity function per se, or, in current
nomenclative, they are not with the general, subhalo mass \--- stellar mass relation.
We will attempt to show in this paper that this unexpectedly large gap is also found
in current simulations of galaxy formation such as EAGLE 
\citep{2015MNRAS.446..521S}, Illustris \citep{2014MNRAS.445..175G}, and
our own work \citep{2017ApJ...844...31C}, that it is probably not due to feedback and most likely arises
from gravitationally induced merging processes in groups and clusters and 
to some extent from tidal stripping of gas from satellite systems. A possible explanation of the 
physical basis for the effects was proposed in \citet{1977ApJ...217L.125O}:
merging among bright galaxies makes the first brightest galaxy brighter
(and with less variance) and makes the (new) second brightest galaxy 
fainter. These gravitational processes increase the ratio defined by the 
\citet{1977ApJ...212..311T} parameter,
\beq
t_{1}  \equiv  \frac{\sigma(M_{1})}{\langle  M_{12}\rangle},
\label{eq:t1}
\eeq
which compares the variance in the brightness of the first brightest
galaxy $\sigma(M_1)$, to the mean gap between first and second
brightest systems $M_{12}$ helping to explain ``too big to fail'' and systems
such as the ``lonely giant'', M94 group. A very careful recent study
of the gap statistics by \citet{2017MNRAS.471.2022T} presents
a review of recent statistical studies and their implications.

We will show that modern data confirm the observational data
presented in \citet{1977ApJ...212..311T} from 
\citet{1973ApJ...183..743S}, that $\Lambda$CDM simulations
show the same large $\delm12$ gap and that it is likely due to
gravitational effects, since it also appears in dark matter only sims
and is not altered by changes in feedback physics 
\citep[c.f.][]{2013MNRAS.433.3539G,2016MNRAS.457.1931S}.

But there is one additional effect. The dimensionless quantity $t_1$,
statistically expected \citep[c.f.][]{1977ApJ...212..311T} to be greater
than unity, is even smaller in dark matter only simulations than it is in those
including baryonic physics. And the explanation for this is partly
due to definitions, rather than physics, in hierarchical cosmologies.
When subunits (e.g. subhalos) merge, the material stripped off
the satellite systems is summed up and included in our  {\it definition}
of the parent halo, thus increasing the gap between the parent
and the largest subunit. This effect is less extreme for the stellar
than the dark matter component, since tidal stripping is strongest
for the latter subunits.

In section~\ref{sec:history} we remind readers of the conclusions of the two 1977 
papers quoted earlier concerning apparently anomalous  gaps
found in galaxy group statistics, and present an update of the observational
results. In section~\ref{sec:cosmo} we analyze current simulations
both with and without baryonic physics and in Section~\ref{sec:summary} we present our
conclusions. 

%********************************************************
% Chapter 2. 
%********************************************************
\section{Abbreviated History of the Observed Gaps}\label{sec:history}
The overall luminosity function of galaxies fits well to the
\citet{1976ApJ...203..297S} function which, at the bright end, is roughly
exponential. Picking randomly from that distribution one could populate
synthetic galaxy groups and clusters and check if the resulting 
distributions matched observations. The resulting comparison would show
dramatic failure even though -- by construction -- the total luminosity
function would match the total observed luminosity function. Observed first brightest 
galaxies would be too massive compared to expectations, with the deviation
from expectations greatest in the smallest groups and the variance amongst
groups would be less than expected. Stated differently the 
zeroth order expectation would be that the brightest system had a mass
proportional to the logarithm of the total mass of the group or cluster, but the
variance in first brightest galaxy luminosities is significantly less than
predicted. A dynamical explanation was proposed in 
\citet{1977ApJ...217L.125O} in a very simplified treatment that has been 
confirmed by detailed work done subsequently \citep[e.g.][]{2013MNRAS.435..901L,
2015MNRAS.447.1491L,2015MNRAS.453.4444Z,2018ApJ...860....2G}. The first brightest
system will grow via mergers and the effect is greatest in the smallest 
systems, since at fixed density the merger time scales inversely with the
total cluster mass. This tends to balance the statistical expectation of
more massive first brightest galaxies in more massive systems and 
produces a smaller variance $\sigma(M_1)$ in the magnitude of first brightest 
galaxies than was expected.

But the same merger process is most likely to consume the second
brightest galaxy increasing the gap between the now brighter first
brightest and the now fainter second brightest system. On average
the gap $\delm12$ would thus grow as mergers proceeded and that
growth was demonstrated in \citet{1977ApJ...217L.125O}  quantitatively.

\citet{1973ApJ...183..743S} commented that ``The brighter the dominant
galaxy becomes, the absolutely fainter will be the second and third
ranked members. The rich are rich at the expense of the poor, progressively.''

\citet{1977ApJ...212..311T} invented the ingenious statistic, $t_{1}$ 
(Equation~\ref{eq:t1}), which quantified both changes described above and
then showed mathematically that for galaxies picked randomly from general
distribution functions the quantity $t_1$ would be expected to be greater than
unity. However, when they compared expectations with reality, using the 
data compiled by \citet{1973ApJ...183..743S}, they found the opposite to
be true. In general $t_1 < 1$ and discrepancy was greatest in the smallest
groups.

As early as \citet{1978ApJ...222...23D}, it was pointed out that ``the statistical
model, regardless of the form of the luminosity function cannot fulfill all 
requirements, hence a special process model seems required.'' He based
his conclusion on the magnitudes of the  $\delm12$, gap, the small value
of $\sigma(M_1)$ and the weak correlation between $m_1$ and cluster richness.

\citet{1977ApJ...212..311T}, basing their analysis on the \citet{1973ApJ...183..743S}
cluster data, looking at groups with over 30 members found a variance in the {\it V}
magnitude of first brightest systems which was only $0.035 \pm 0.002$ and a 
value for $t_1$ for these same systems $t_1 = 0.55 \pm 0.13$ far below the
expectation (given the luminosity function) of $t_1 > 1$. For Gamma function, 
Schechter function and even double exponential functions the expected value
is $t_1 \sim 1.3$. \cite{2006MNRAS.366..373L} examined 2099 $\rm deg^2$ of
Sloan Digital Sky Survey (SDSS) data again looking for bright galaxies and 
searching in the redshift range
$0.12 < z < 0.38$ using the $r$ band best for detecting luminous red galaxies.
Again they found large gaps between first and second brightest systems with a
characteristic value of $ \langle   M_{12} \rangle \sim 0.8$ mag, very similar to the value obtained by
\citet{1977ApJ...212..311T}, from the \citet{1973ApJ...183..743S} data.

\citet{2006MNRAS.366..373L} data give a value for $t_1 = 0.75 \pm 0.12$ for 
the richer clusters and $t_1 =0.27 \pm 0.06$ for the poorer systems consistent  
with \citet{1977ApJ...212..311T} and grossly inconsistent with the statistical
expectation of $t_1 \sim 1.3$. They again formed a gap of  
$\langle M_{12} \rangle \sim 0.87$ magnitude. Clearly the observed gaps are far bigger than
what we would have expected from the luminosity function, i.e., intermediate 
mass galaxies are missing, and the first brightest systems are more standardized 
than expected.

There is a very relevant later paper by \cite{2014ApJ...782...23S} entitled ``The 
Statistical Nature of the Brightest Group Galaxies'' which examines the problem 
from a different angle. They also compute the Tremaine-Richstone statistic and 
again find $t_1$ significantly less than unity for the large sample of groups that 
they study with typical observed values being $0.70 \pm 0.05$. They attribute 
this to their finding that the first brightest galaxies are `too bright' and, when they 
correct down the brightness of these systems, they conclude that the gaps are 
close to expectations. But there is however a bit of circular reasoning involved 
in this explanation. They use the total luminosity of the systems to estimate the 
halo masses and then ask what is the expected luminosity of the first brightest 
system given that halo mass. But of course if the satellite systems are too faint, 
then the halo mass is underestimated and then the `expected' luminosity of the 
first brightest system is found to be low and the observed BCG is consequently 
`too bright'. The Tremaine-Richstone criterion itself is not subject to this criticism. 
So, in sum, the \cite{2014ApJ...782...23S} paper agrees that the observed gaps 
are larger than statistically expected but can not make a clean argument as to 
how much of this is due to the BCG being brighter than expected or to the 
satellites being fainter than expected.

In a related paper \cite{2010ApJ...715.1486L} studying more massive groups 
had found the value $t_1 = 0.93 \pm 0.01$, but do not draw a firm conclusion 
as to the origin of the statistical anomaly.

Notice that we have not discussed galaxy formation, feedback or any of the
factors that determine the halo mass -- galaxy mass relation. All of the 
discussion concerns the expectations of the properties of the two most
massive systems in a group, {\it given} the luminosity (or mass) function. Thus,
the too big gap -- which can contribute to the
``too big to fail'' problems must be understood quite separately from the 
processes such as efficiency of star formation, feedback etc, that determine
the overall luminosity function.

\begin{table*}
   \begin{center}
   \caption{EAGLE simulation}
    {
   \begin{tabular}{c|c|c|c|c|c|c|c}\hline\hline \label{tab:eagle1}
galaxy number  & $\langle m_{\rm halo,1} \rangle$\tablenotemark{a}   & $\langle M_1 \rangle$  &    $ \sigma(M_1)$   &    $\langle   M_{12} \rangle$ &    $\sigma(M_{12})$    &     $t_1$     &    Number of groups\\
  \hline
12-24 &  $9.09 \times 10^{12}$   &     -23.55  &        0.54   &         1.89       &     1.13     &       0.28      &      152\cr
25-49  &  $2.07 \times 10^{13}$  &     -24.26  &        0.43   &         1.57       &     1.00     &       0.27      &      73\cr
50-74   & $4.17 \times 10^{13}$   &    -24.84  &        0.28   &         1.92       &     0.70     &       0.15      &      19\cr
75-150  & $6.98 \times 10^{13}$   &   -25.26  &        0.37   &         1.60       &     0.69     &       0.23      &      12\cr
\hline
Mean &   $1.77 \times 10^{13}$  &   -23.93   & 0.48  &  1.79  &  1.04   & $0.27 \pm 0.03$ &  256\cr
  \hline\hline
   \end{tabular}}
   \end{center}
       \begin{flushleft}
       $^{a}$ Average halo mass of the first brightest galaxies in solar mass.
   \end{flushleft}
 \end{table*}

  \begin{table*}
   \begin{center}
   \caption{EAGLE `randomized' simulation}
    {
   \begin{tabular}{c|c|c|c|c|c|c|c}\hline\hline \label{tab:eagle2}
galaxy number  &  $\langle m_{\rm halo,1} \rangle$     &     $\langle M_1 \rangle$  &    $ \sigma(M_1)$   &    $\langle   M_{12} \rangle$ &    $\sigma( M_{12})$    &     $t_1$      &    Number of groups\\
  \hline
12-24   &  $3.27 \times 10^{12}$   &  -22.01    &      1.33     &       1.22     &       1.04      &      1.09     &       152\cr
25-49    & $4.35\times 10^{12}$    &  -22.52    &      1.16     &       0.79     &       0.69       &     1.46    &        73\cr
50-74   &    $1.08\times 10^{13}$  &  -22.40  &        0.82     &       0.89     &       0.54   &         0.92    &        19\cr
75-150   &$9.91\times 10^{12}$    &  -23.39  &        0.77     &       0.73     &       0.46    &        1.06     &       12\cr
\hline
Mean &   $4.45\times 10^{13}$  &    -22.38  &  1.15 &   1.10 &   0.86  & $1.06\pm 0.06$ &  256\cr
  \hline\hline
   \end{tabular}}
   \end{center}
 \end{table*}

  \begin{figure}
 \includegraphics[width=\columnwidth]{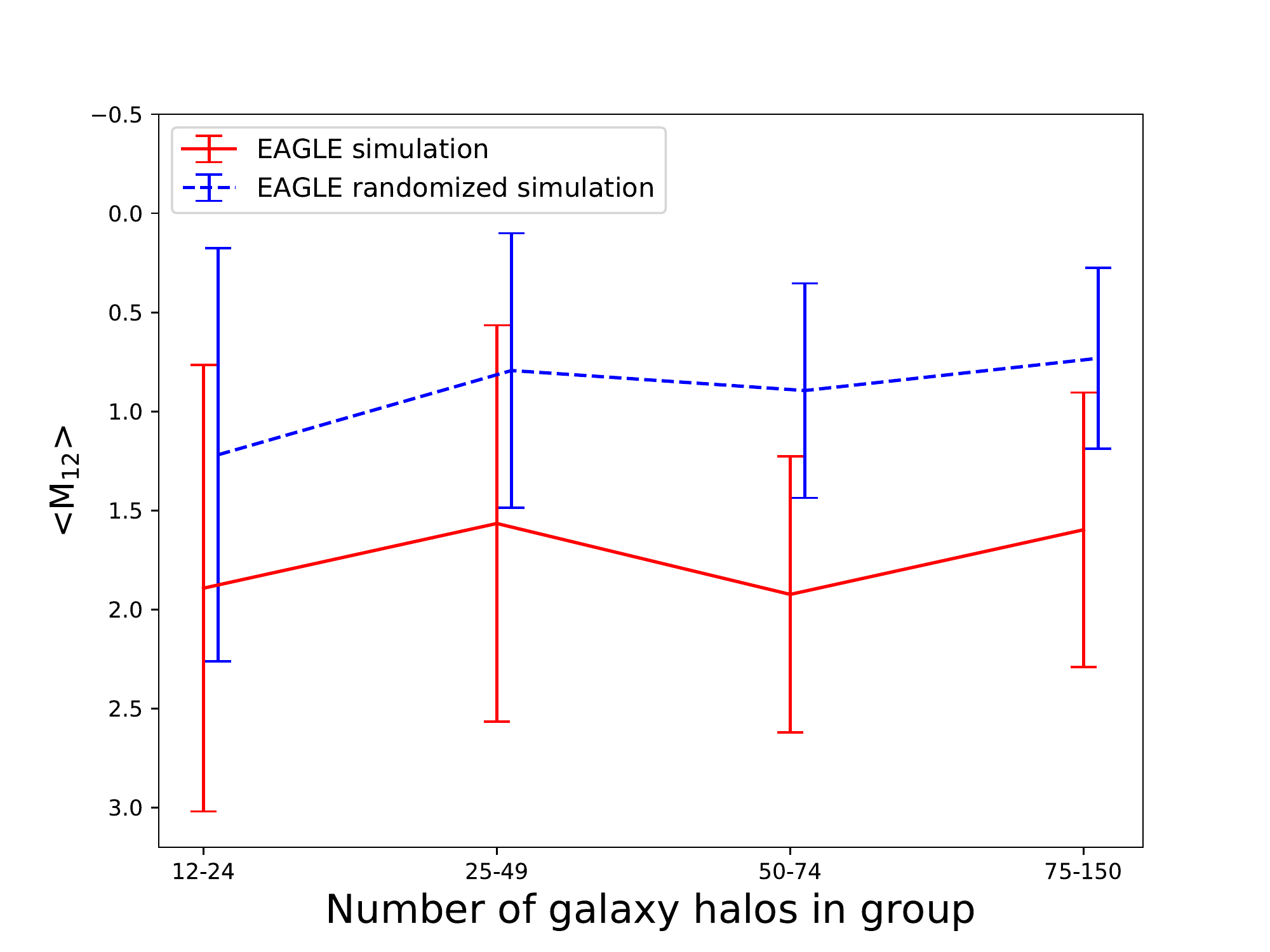}
 \caption{$M_{12}$ presented in EAGLE simulation (red) and in the `randomized' data (blue).
 \label{fig:eagle1} }
\end{figure}

\begin{figure}
 \includegraphics[width=\columnwidth]{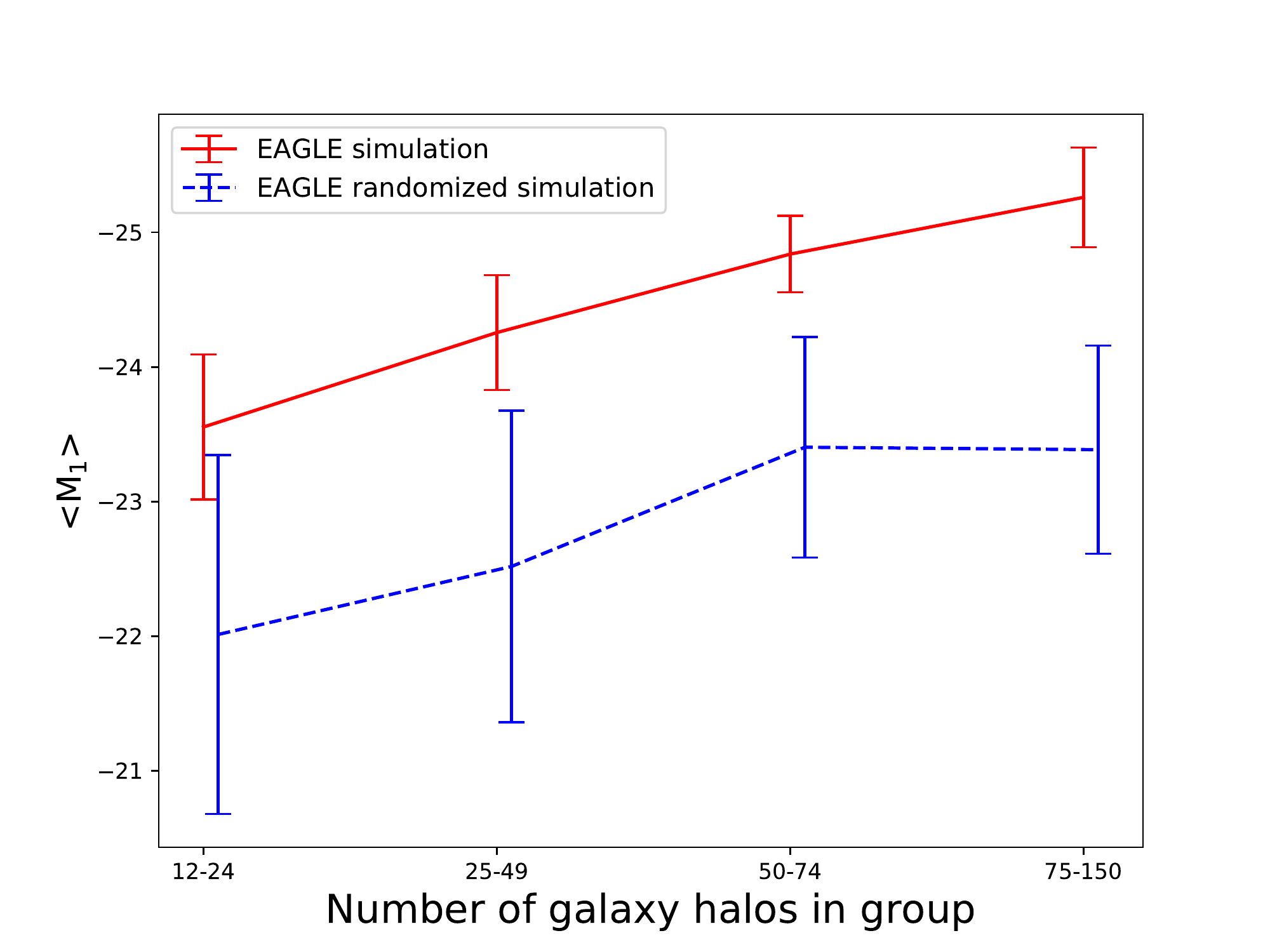}
 \caption{Average first-ranked {\it r} magnitude as a function of number of galaxy members in EAGLE simulation.
 \label{fig:eagle2} }
\end{figure}

\section{Cosmological simulations}\label{sec:cosmo}
These anomalies in the real world were observed long before there were
detailed, physically based numerical simulations of galaxy formation with
which to compare the strange results. The papers which have pointed out
the ``too big to fail'' problem have not \--- so far as we are aware \--- compared
expectations with cosmological simulations. Rather, they have asked, given the brightest
galaxy in a group and its expected dark matter halo, what are the expected
lower mass halos in the group and what galaxies are expected to live within
them. For this they use the \cite{1976ApJ...203..297S} function giving the average mass  
distribution of dark matter halos on an equivalent statistical model. And the
authors universally find that many intermediate mass galaxies are expected
in the groups which are not there at a statistically significant level: the missing
systems are ``too big to fail''.

Let us see what the simulations tell us. We have looked at three sets: EAGLE 
simulation \citep{2015MNRAS.446..521S}, IllustrisTNG simulation 
\citep{2018MNRAS.473.4077P,2018MNRAS.475..676S}, and our own 
\citep{2017ApJ...844...31C}. The 
first two sets of simulations have had enough statistical power to determine if
they can match the observed luminosity function for bright galaxies and numerous 
papers \citep[e.g.][]{2014MNRAS.444.1518V,2015MNRAS.450.1937C} detail 
their success. So, whether the CDM paradigm is right or wrong, the physical
processes that they implement give them a luminosity function, above and below
the $L_{\ast}$ break, which matches real data.

We have looked at the publicly available output from these groups to see if their
results do or do not match observations on statistical expectations with regard to the 
\cite{1977ApJ...212..311T} $t_1$ statistics. We use $r$-band magnitudes throughout
this study.

\subsection{EAGLE simulation}\label{sec:eagle}
The EAGLE simulation is a publicly available \citep{2016A&C....15...72M} 
suite of cosmological hydrodynamical simulation 
\citep{2015MNRAS.446..521S,2015MNRAS.450.1937C}
It assumes a standard $\Lambda$ cold dark matter cosmology from
the Planck-1 \citep{2014A&A...571A..16P}, $\rm \Omega_m = 1 - \Omega_{\Lambda}=0.307$,
$\rm \Omega_b = 0.04825$, $h=0.6777$, $\sigma_8 = 0.8288$, and $n_{\rm S} =0.9611$.
The simulation suite is run with a modified version of the GADGET-3 $N$-body Tree-PM
smoothed particle hydrodynamics (SPH) code \citep{2005MNRAS.364.1105S},
and includes an updated  formulation of SPH, new time stepping, and 
sub-grid physics \citep[see][for details]{2015MNRAS.446..521S}. 
In this study, we use RefL0100, which has a cosmological volume 
of $(100\,\mathrm{comoving\,Mpc})^3$ and a baryonic mass 
resolution of $1.81\times 10^6 \Msun$. We refer the readers to the
introductory papers \citep{2015MNRAS.446..521S,2015MNRAS.450.1937C} 
for a complete descriptions of sub-grid physics models.

\citet{}
\begin{table*}
   \begin{center}
   \caption{IllustrisTNG  simulation}
    {
   \begin{tabular}{c|c|c|c|c|c|c|c}\hline\hline \label{tab:illus1}
galaxy number  &  $\langle m_{\rm halo,1} \rangle$ &       $\langle M_1 \rangle$  &    $ \sigma(M_1)$   &    $\langle   M_{12} \rangle$ &    $\sigma(M_{12})$    &     $t_1$      &    Number of groups\\
  \hline
12-24    & $8.27 \times 10^{12}$  & -22.47   &    0.47    &     1.55    &      0.97  &  0.30  & 187\cr
25-49   &  $ 1.95\times 10^{13}$  & -23.14  &     0.47    &     1.60   &       0.91  &  0.30  &    82\cr
50-74     & $ 3.87\times 10^{13}$ & -23.65  &     0.37    &     1.19   &       0.75  &  0.31   &    23\cr
75-149   & $ 6.60\times 10^{13}$ &  -24.00   &    0.44   &      1.47    &      0.98  &  0.30  &    26\cr
150$>$  & $ 1.85\times 10^{14}$ &  -24.76   &    0.44    &     1.04     &     0.81   & 0.42  &    14\cr
\hline
Mean &   $ 2.63 \times 10^{13}$ &  -22.93   & 0.46    &1.51  &  0.93  &  $0.31\pm 0.02$ &  332\cr
  \hline\hline
   \end{tabular}}
   \end{center}
 \end{table*}
 
 \citet{}
\begin{table*}
   \begin{center}
   \caption{IllustrisTNG  `randomized' simulation}
    {
   \begin{tabular}{c|c|c|c|c|c|c|c}\hline\hline \label{tab:illus2}
   %  1.86655e+12  4.18305e+12  7.98007e+12  1.45676e+13  6.59287e+13  7.79244e+12
galaxy number  &  $\langle m_{\rm halo,1} \rangle$ &       $\langle M_1 \rangle$  &    $ \sigma(M_1)$   &    $\langle   M_{12} \rangle$ &    $\sigma(M_{12})$    &     $t_1$      &    Number of groups\\
  \hline
12-24   &  $1.87 \times 10^{12}$ & -21.77    &  1.29    &    1.16   &      0.95 &  1.12 &     187\cr
25-49    &$ 4.18 \times 10^{12}$  & -22.61    &  0.98    &    0.94    &     0.73  & 1.04  &     82\cr
50-74   & $ 7.98 \times 10^{12}$  & -23.19    &  1.00    &    0.73    &     0.74   &1.36   &    23\cr
75-149  & $ 1.46 \times 10^{13}$ &  -23.47   &   0.93   &     0.59    &     0.70  & 1.58  &     26\cr
150$>$ & $  6.59 \times 10^{13}$  & -23.84  &    0.65   &     0.45    &     0.34 &  1.45  &     14\cr
\hline
Mean   & $ 7.79 \times 10^{12}$    &  -22.30   & 1.14   & 1.00  &  0.84  &  $1.17\pm 0.16$&  332\cr
  \hline\hline
   \end{tabular}}
   \end{center}
 \end{table*}
 
 \begin{figure}
 \includegraphics[width=\columnwidth]{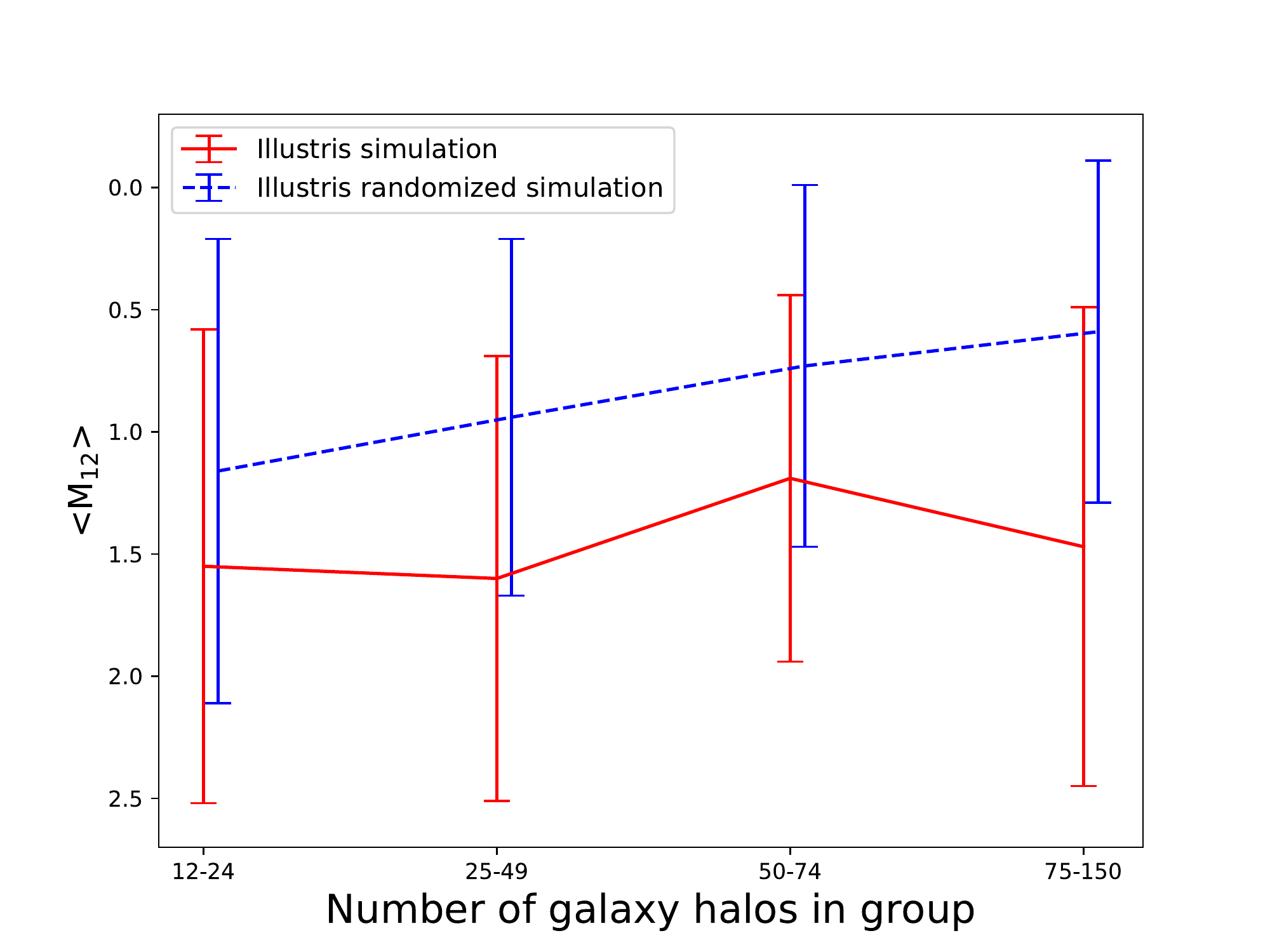}
 \caption{$M_{12}$ presented in IllustrisTNG simulation (red) and in the `randomized' data (blue).
 \label{fig:illus1} }
\end{figure}

\begin{figure}
 \includegraphics[width=\columnwidth]{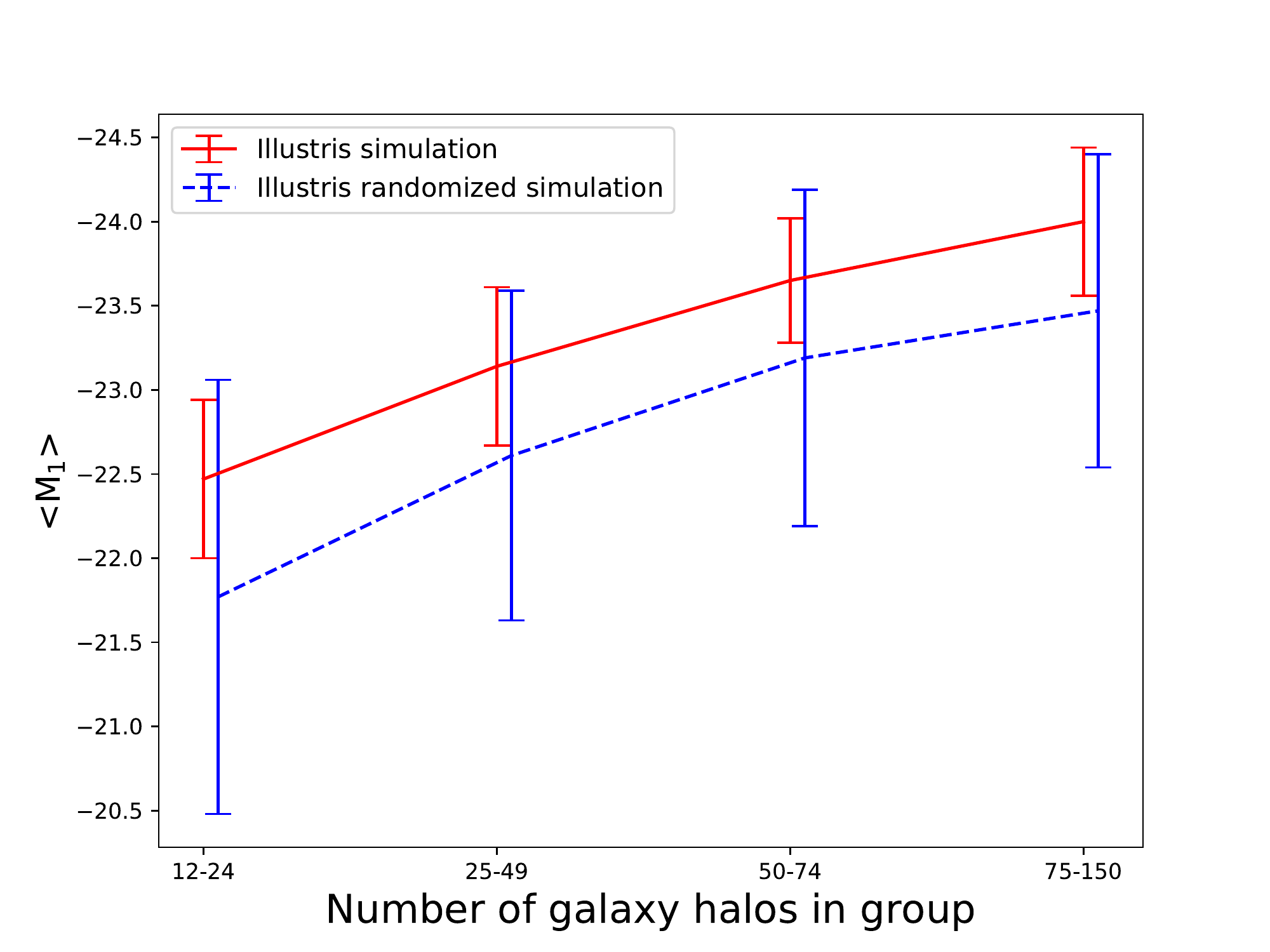}
 \caption{Average first-ranked {\it r} magnitude as a function of number of galaxy members in IllustrisTNG simulation.
 \label{fig:illus2} }
\end{figure}

We bin EAGLE simulation results into different size galaxy groups
from smallest (12--24 objects) with stellar mass greater than 
$10^8 \Msun$ to more massive systems with 75--150 galaxies 
(See Table~\ref{tab:eagle1}). The red lines in Figures~\ref{fig:eagle1} 
and \ref{fig:eagle2} show the average magnitude gap between 
first-brightest and second-brightest galaxies, $\langle M_{12} \rangle$,
in each group and brightness of the first brightest galaxy 
$\langle M_1 \rangle$. Overall, simulated galaxies in EAGLE
show $\sim 1.7$ magnitude gap between first and second brightest
galaxies, which is similar to the gap reported by
\citet{2007MNRAS.376..841V} with 2-degree Field Galaxy Redshift
Survey data.

Next, we put all groups in one bin and 
randomly repopulate each group from the collective bin, keeping 
the same number of galaxies, in each galaxy group or cluster 
(See Table~\ref{tab:eagle2}). By construction, this keeps the overall 
luminosity function invariant. Then we recompute 
$\langle M_{12} \rangle$,  $\langle M_1 \rangle$ and $ \sigma(M_1)$
for each of Figure~\ref{fig:eagle1} and \ref{fig:eagle2}.

Note how the gap is systematically larger in the original 
($\langle M_{12} \rangle=1.79$) than in the randomized groups 
($\langle M_{12} \rangle=1.10$) and the first brightest galaxy is 
brighter by over a magnitudes in each original set of galaxy groups 
than in the randomized versions of the same objects. Somehow in 
each system the dominant member ``knows'' it is first and becomes
more dominant. 

In Tables~\ref{tab:eagle1} and \ref{tab:eagle2}, we summarize 
these results and compute the Tremaine-Richstone parameter 
$t_1$, statistic via Equation~\ref{eq:t1}. We note that the simulated 
data in Table~\ref{tab:eagle1} shows very low values of $t_1$, 
typically around 1/4, even lower than the real data analyzed by 
\cite{1977ApJ...212..311T} and \cite{2006MNRAS.366..373L}. So 
this ``anomalous'' behavior is reproduced by the simulations
which have larger than expected gaps and consequently could be 
accused of not having the expected second brightest galaxies.

Then Table~\ref{tab:eagle2} shows the $t_1$ statistic for the 
randomized data and -- lo and behold -- it exactly follows the 
statistical expectations with $t_1 \approx 1.06 \pm 0.06$.

\subsection{IllustrisTNG simulation}\label{sec:illustris}
The IllustrisTNG simulation \citep{2018MNRAS.473.4077P,
2018MNRAS.475..624N,2018MNRAS.475..676S}
is a publicly available  suite of cosmological simulation 
\citep{2015A&C....13...12N}, an extension of the Illustris 
simulation \citep{2014MNRAS.445..175G,2014MNRAS.444.1518V,
2015MNRAS.452..575S}. It adopts the Planck Collaboration XIII 
cosmological parameters \citep{2016A&A...594A..13P}, with 
$\rm \Omega_m = 1 - \Omega_{\Lambda}=0.3089$,
$\rm \Omega_b = 0.0486$, $h=0.6774$, $\sigma_8 = 0.8159$, 
and $n_{\rm S} =0.9667$. The simulation is evolved with the 
AREPO moving-mesh code \citep{2010MNRAS.401..791S}, and 
employs a number of improvements of sub-grid physics models for
stellar and AGN feedback, and black hole growth 
\citep{2018MNRAS.475..648P,2018MNRAS.479.4056W}. The 
adopted fiducial simulation we use in this paper (TNG-100) has a 
cosmological volume of $(110.7~\rm{Mpc} )^{3}$ and a baryonic 
mass resolution of $1.4 \times 10^{6} \Msun$. We refer the readers
to the introductory papers of original Illustris 
\citep[e.g.][]{2014MNRAS.445..175G} and IllustrisTNG 
\citep[e.g.][]{2018MNRAS.473.4077P} for further details.

Now we repeat the same exercise done previously with the EAGLE 
simulations using now the IllustrisTNG simulations, for the galaxies 
with stellar mass greater than $10^8 \Msun$. The results are 
shown in Figure~\ref{fig:illus1} and \ref{fig:illus2} and summarized 
in Table~\ref{tab:illus1} and \ref{tab:illus2}. Again the published 
simulations show ``too big'' a gap and ``too bright'' first brightest 
galaxies and -- correspondingly -- the $t_1$ statistic is smaller 
($t_1 = 0.31 \pm 0.02$) than statistically expected. In the 
randomized data $t_1$ is again much higher 
($t_1 = 1.17 \pm 0.16$) and larger than unity. The gap is again 
larger by $0.51$ in the initial fiducial data than in the randomized
data.

What is the cause of these fascinating results? We mentioned in 
the Introduction several possible physical effects that could do it: 
tidal stripping of satellites, feedback from the central galaxies 
removing cold gas from the environments of satellite systems, 
ram pressure stripping and finally merging. One could imagine 
complicated simulation tests where each of these effects was
switched on or off to determine its consequences for the statistical 
tests, but there is a far simpler approach that can be taken. All 
these -- and many other ``baryonic'' effects are missing in the 
preliminary dark matter only simulations that the EAGLE and 
Illustris group have performed. We will discuss this in 
Section~\ref{sec:dmonly}.

\begin{table*}
   \begin{center}
   \caption{EAGLE Dark Matter Only Simulation}
    {
   \begin{tabular}{c|c|c|c|c|c|c|c}\hline\hline \label{tab:dm1}
galaxy number &          $\langle m_{\rm halo,1} \rangle$  &     $\langle M_1  \rangle ^a$  &    $ \sigma(M_1)$   &    $\langle   M_{12} \rangle$ &    $\sigma(M_{12})$    &     $t_1$     &    Number of groups\\
  \hline
12-24    & $1.09 \times 10^{12}$ &     -25.92   &       0.66     &       3.43     &       1.44    &        0.19    &        1111 \cr
25-49    &  $2.60 \times 10^{12}$  &   -26.91    &      0.55     &       3.46     &       1.44    &        0.16    &        485 \cr
50-74    & $4.66 \times 10^{12}$    &  -27.57    &      0.50     &       3.30     &       1.43    &        0.15     &       172 \cr
75-150  & $ 8.33 \times 10^{12}$   &    -28.21   &       0.49    &        3.43    &       1.43    &        0.14     &       181 \cr
\hline
Mean &  $2.45 \times 10^{12}$  &   -26.52  &  0.60  &  3.43   & 1.44  &  $0.17\pm 0.02$ & 1949 \cr
  \hline\hline
   \end{tabular}}
   \end{center}
      \begin{flushleft}
    $^{a}$ Note: we define `mass-magnitude' of dark matter halo mass as $M = - 2.5 \rm log (m_{\rm halo} / \Msun$) + 4
   \end{flushleft}  
 \end{table*}

 \begin{table*}
   \begin{center}
   \caption{EAGLE Dark Matter Only Simulation `Randomized'}
    {
   \begin{tabular}{c|c|c|c|c|c|c|c}\hline\hline \label{tab:dm2}
galaxy number &   $\langle m_{\rm halo,1} \rangle$ &    $\langle M_1  \rangle ^a$  &    $ \sigma(M_1)$   &    $\langle   M_{12} \rangle$ &    $\sigma(M_{12})$    &     $t_1$     &    Number of groups\\
  \hline
12-24    & $ 2.43\times 10^{11}$   &   -23.08    &      1.44     &       1.15   &         1.13    &        1.24   &         1111\cr
25-49    &  $ 4.73\times 10^{11}$  &   -24.96   &       1.43     &       1.18  &          1.11     &       1.21 &           485\cr
50-74    & $ 7.55\times 10^{11}$   &   -24.53    &      1.41     &       1.11   &         1.09   &         1.27  &          172\cr
75-150  &  $ 1.49\times 10^{12}$  &   -25.27     &     1.35      &      1.12     &       1.04     &       1.20   &         181\cr
\hline
Mean &   $ 4.60\times 10^{11}$ &    -23.77 &   1.58   & 1.34   & 1.24  &  $1.18\pm 0.06$ & 1949 \cr
  \hline\hline
   \end{tabular}}
         \begin{flushleft}
    $^{a}$ Note: we define `mass-magnitude' of dark matter halo mass as $M = - 2.5 \rm log (m_{\rm halo} / \Msun$) + 4
   \end{flushleft}  
   \end{center}
 \end{table*}

 \begin{table*}
   \begin{center}
   \caption{IllustrisTNG Dark Matter Only Simulation}
    {
   \begin{tabular}{c|c|c|c|c|c|c|c}\hline\hline \label{tab:dm3}
galaxy number &   $\langle m_{\rm halo,1} \rangle$ &    $\langle M_1  \rangle ^a$  &    $ \sigma(M_1)$   &    $\langle   M_{12} \rangle$ &    $\sigma(M_{12})$    &     $t_1$     &    Number of groups\\
  \hline
12-24   & $ 1.86 \times 10^{12}$   &  -25.69  &    0.67   &     3.36      &   1.36 &  0.20 &   840 \cr
25-49   &  $ 4.18 \times 10^{12}$ &  -26.63   &   0.56   &     3.27   &     1.37  &  0.17   &  388 \cr
50-74   & $ 7.98 \times 10^{12}$  & -27.37  &    0.44   &     3.47   &      1.19  & 0.13 &     145 \cr
75-149  & $ 1.46  \times 10^{13}$  &  -28.03 &     0.42    &    3.49    &     1.34  & 0.12  &     118 \cr
150$>$ & $ 6.59 \times 10^{13}$ &  -29.33   &   0.88     &  2.95    &     1.40   & 0.30  &    106 \cr
\hline
 Mean & $ 8.17 \times 10^{12}$ & -26.49   & 0.62  &  3.33    &1.35  &  0.19    $\pm$ 0.04 & 1597\cr
  \hline\hline
   \end{tabular}}
   \end{center}
      \begin{flushleft}
    $^{a}$ Note: we define `mass-magnitude' of dark matter halo mass as $M = - 2.5 \rm log (m_{\rm halo} / \Msun$) + 4
   \end{flushleft}  
 \end{table*}
 
  \begin{table*}
   \begin{center}
   \caption{IllustrisTNG Dark Matter Only Simulation `Randomized'}
    {
%  2.50148e+12  3.47490e+12  6.83245e+12  1.37544e+13  3.59333e+13  6.18172e+12
   \begin{tabular}{c|c|c|c|c|c|c|c}\hline\hline \label{tab:dm4}
galaxy number &    $\langle m_{\rm halo,1} \rangle$ &  $\langle M_1  \rangle ^a$  &    $ \sigma(M_1)$   &    $\langle   M_{12} \rangle$ &    $\sigma(M_{12})$    &     $t_1$     &    Number of groups\\
  \hline
12-24   & $  2.50\times 10^{12}$ &    -23.45  &    2.10   &     1.71    &     1.66  & 1.23  &   840\cr
25-49   & $  3.47 \times 10^{12}$ &   -24.92  &    2.06   &     1.89    &     1.70  & 1.09  &  388\cr
50-74   & $  6.83 \times 10^{12}$ &   -25.67  &    1.80   &     1.59    &     1.46  & 1.14  &  145\cr
75-149  & $  1.38 \times 10^{13}$ &   -26.58   &   1.61 &       1.49    &     1.20  & 1.08  &  118\cr
150$>$ & $  3.59\times 10^{13}$&   -28.24  &    1.49   &     1.06    &     1.00 &  1.40 &  106\cr
\hline
 Mean & $  6.18 \times 10^{12}$&   -24.56   & 1.99  &  1.68   & 1.57   & 1.19 $\pm$   0.09  &1597\cr
  \hline\hline
   \end{tabular}}
   \end{center}
      \begin{flushleft}
    $^{a}$ Note: we define `mass-magnitude' of dark matter halo mass as $M = - 2.5 \rm log (m_{\rm halo} / \Msun$) + 4
   \end{flushleft}  
 \end{table*}

  \begin{table*}
   \begin{center}
   \caption{Summary of observation and simulations}
    {
   \begin{tabular}{c|c|c}\hline\hline \label{tab:summary}
 &   $\langle M_{12} \rangle$    & $t_1$ \\
  \hline
\cite{1977ApJ...212..311T} Analysis$^{\ast}$   &        $0.80 \pm 0.71$    & 0.45     \cr
\cite{2006MNRAS.366..373L} Analysis$^{\ast\ast}$   & $0.88 \pm 0.07$& $0.34 \pm 0.02$\cr
\hline
EAGLE $\Lambda$CDM galaxy simulation   & $1.79 \pm 1.04$             & $0.27 \pm 0.03$ \cr
IllustrisTNG $\Lambda$CDM galaxy simulation    & $1.51 \pm 0.93$      & $0.31 \pm 0.02$ \cr
\cite{2017ApJ...844...31C} galaxy simulation & $2.57 \pm 1.00$             & $0.21 \pm 0.14 $\cr
EAGLE $\Lambda$CDM dark matter only simulation & $3.43 \pm 1.44$ & $0.17\pm0.02$ \cr
 IllustrisTNG $\Lambda$CDM dark matter only simulation & $3.33 \pm 1.35$ & $0.19\pm0.04$ \cr 
  \hline\hline
   \end{tabular}}
   \end{center}
         \begin{flushleft}
    $^{\ast}$ \cite{1973ApJ...183..743S} data \\
    $^{\ast\ast}$ SDSS data from \cite{2003AJ....126.2081A}
   \end{flushleft} 
 \end{table*}

 \subsection{Zoom-in simulation of galaxy-group size halos}\label{sec:choi}
\begin{figure}
 \includegraphics[width=\columnwidth]{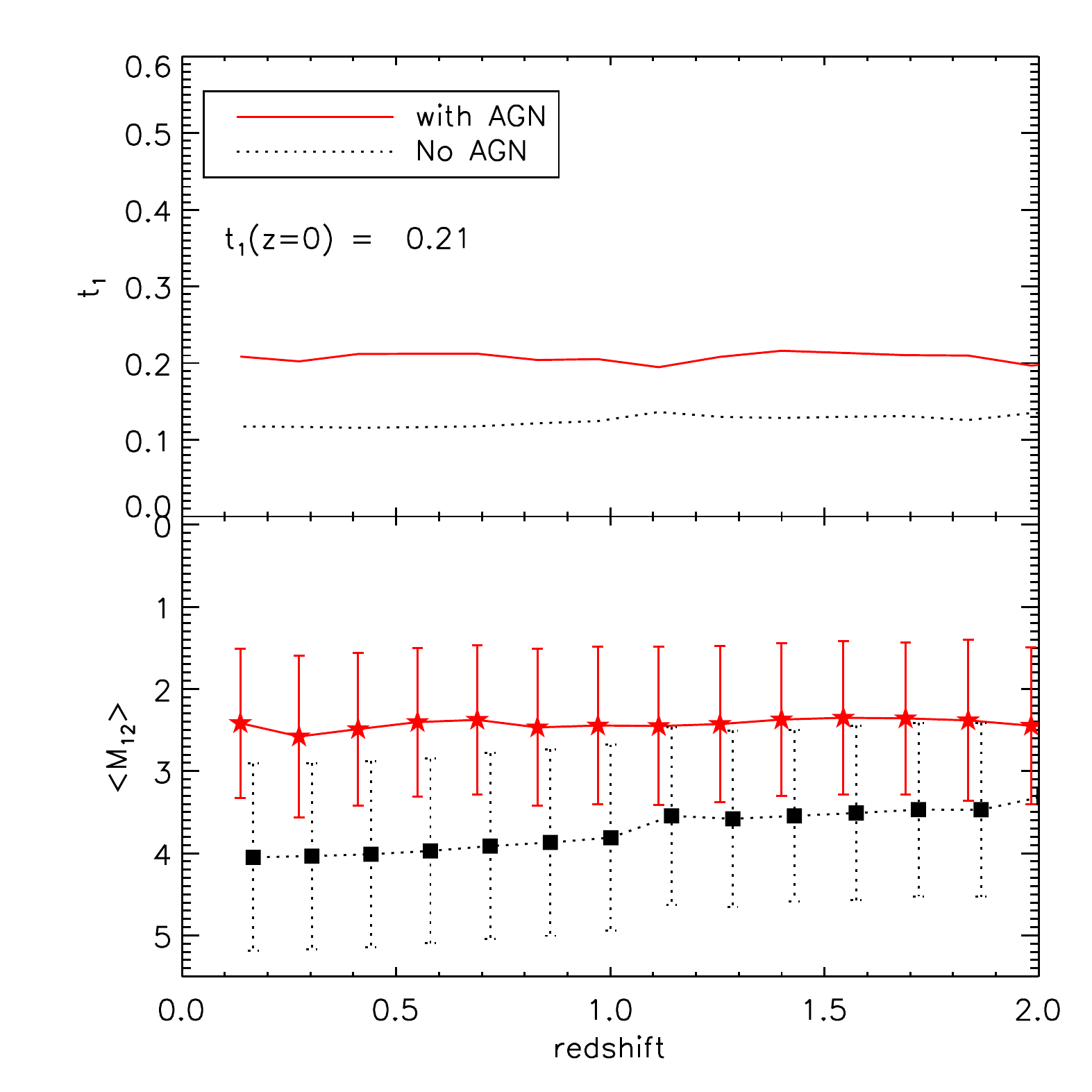}
 \caption{Evolution of $t_{1} $ parameter ($t_{1} = {\sigma(M_{1})}/{\langle  M_{12}\rangle}$) and the magnitude gap $\langle M_{12} \rangle$ in \cite{2017ApJ...844...31C} simulation from $z=2$ to 0.
 \label{fig:zoomevol} }
\end{figure}

This time we study the evolution of the magnitude gap and $t_1$ 
parameter in 30 massive halos with present-day halo masses of 
$1.4 \times 10^{12} \Msun \le M_{\mathrm{vir}} \le 2.3 \times 
10^{13} \Msun$ in cosmological zoom-in hydrodynamic simulations. 
We used two sets of 30 high-resolution zoom-in simulations from 
\cite{2017ApJ...844...31C} simulated {\it with} and {\it without} AGN
feedback. The most massive and brightest central galaxies in these 
zoom-in simulations have stellar masses of $8.2 \times 10^{10} 
\Msun \le M_{\ast} \le 1.5 \times 10^{12} \Msun$ at $z=0$. 

The physics implemented in the simulations 
includes star formation, mechanical supernova 
feedback, wind feedback from massive stars, AGB stars and metal cooling and 
diffusion. The AGN feedback model is adopted from \citep{2012ApJ...754..125C,
2014MNRAS.442..440C} and consists of two main components: (1) mechanical 
feedback via high velocity broad absorption line {\it winds}, which deposits 
energy, mass and momentum into the adjacent gas, and (2) radiative feedback 
from X-ray radiation of the accreting black holes via the photoionization and the 
Compton heating following \cite{2004MNRAS.347..144S}.
The simulation set used in this section is presented in 
\cite{2017ApJ...844...31C}, and we refer the reader to this paper for
further details.

In Figure~\ref{fig:zoomevol}, we show the evolution of $t_1$ parameter 
and the $r$-band magnitude gap between  first- and second- brightest
galaxies $\langle M_{12} \rangle$ from $z=2$ to $z=0$ for two sets of 
zoom-in simulations, run with and without the AGN feedback. 
We have almost constant $t_1$ parameter from $z=2$ to $z=0$, in both
simulations with and without AGN feedback, and a mild increase of 
$\langle M_{12} \rangle$  with time from  $z=2$ to $z=0$ in simulations
without AGN feedback, implying some brightening of the first ranked 
galaxy compared with the second ranked galaxy over time.

Also, as expected AGN activity does tend to strip satellite systems 
\citep{dashyan2019,2019MNRAS.484.2433S} increasing the gap 
and further decreasing $t_1$, but these real effects are not the dominant 
ones. Instead, an excess and prolonged star formation in first-ranked 
galaxies shows a dominant effect, showing an increase in $\langle M_{12} \rangle$
with time.
However, we see that for all three fiducial sets of simulations presented in
\ref{sec:eagle}, \ref{sec:illustris}, and \ref{sec:choi}, the $t_1$ statistic is similar, 
$t_1 \sim 0.3$, and it is even below the value found in the observational
data.

\subsection{Dark matter only simulation}\label{sec:dmonly}
\begin{figure}
 \includegraphics[width=\columnwidth]{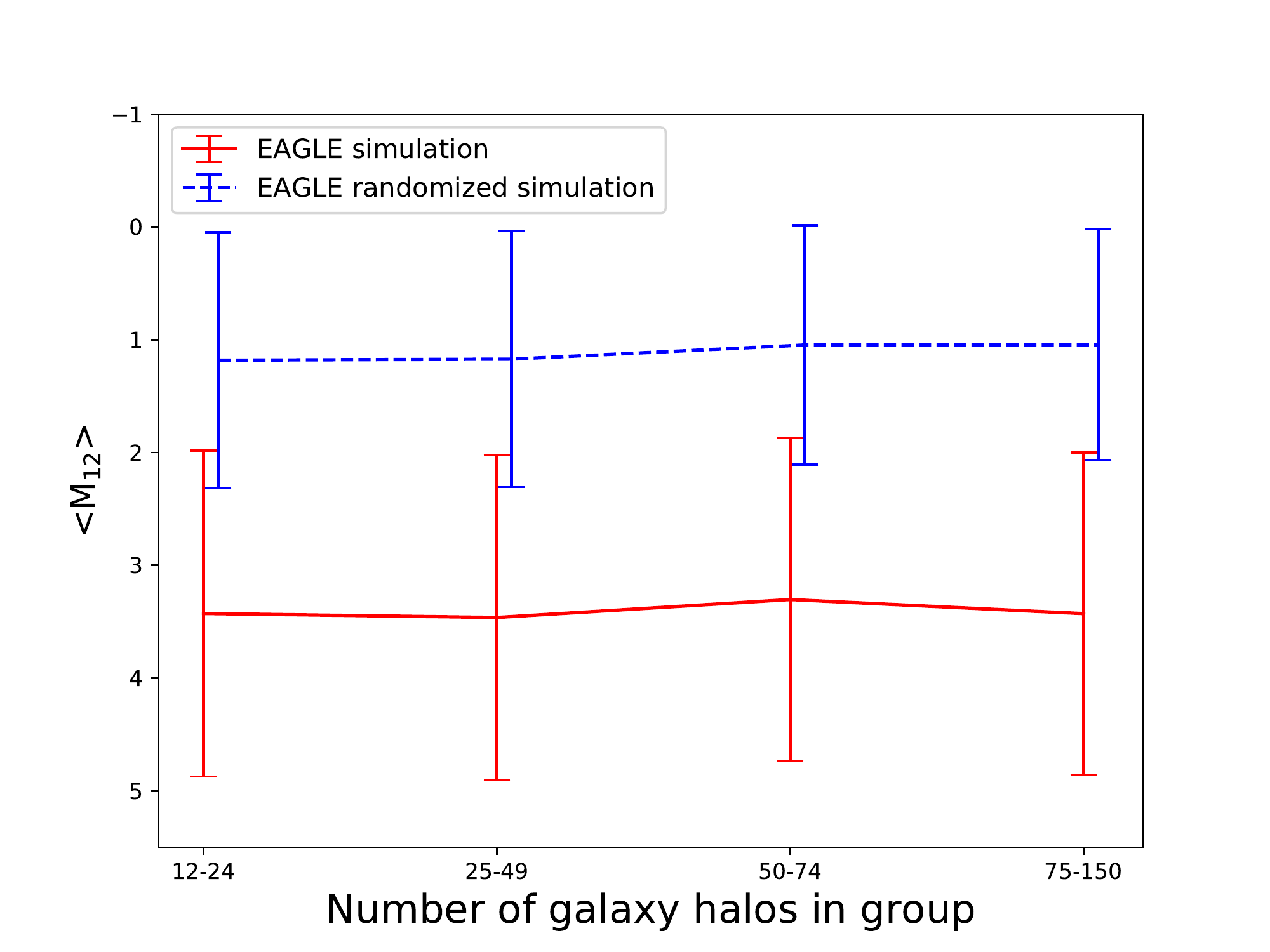}
 \caption{$M_{12}$ presented in EAGLE dark matter only simulation (red) and in the `randomized' data (blue).
 \label{fig:dm1} }
\end{figure}

\begin{figure}
 \includegraphics[width=\columnwidth]{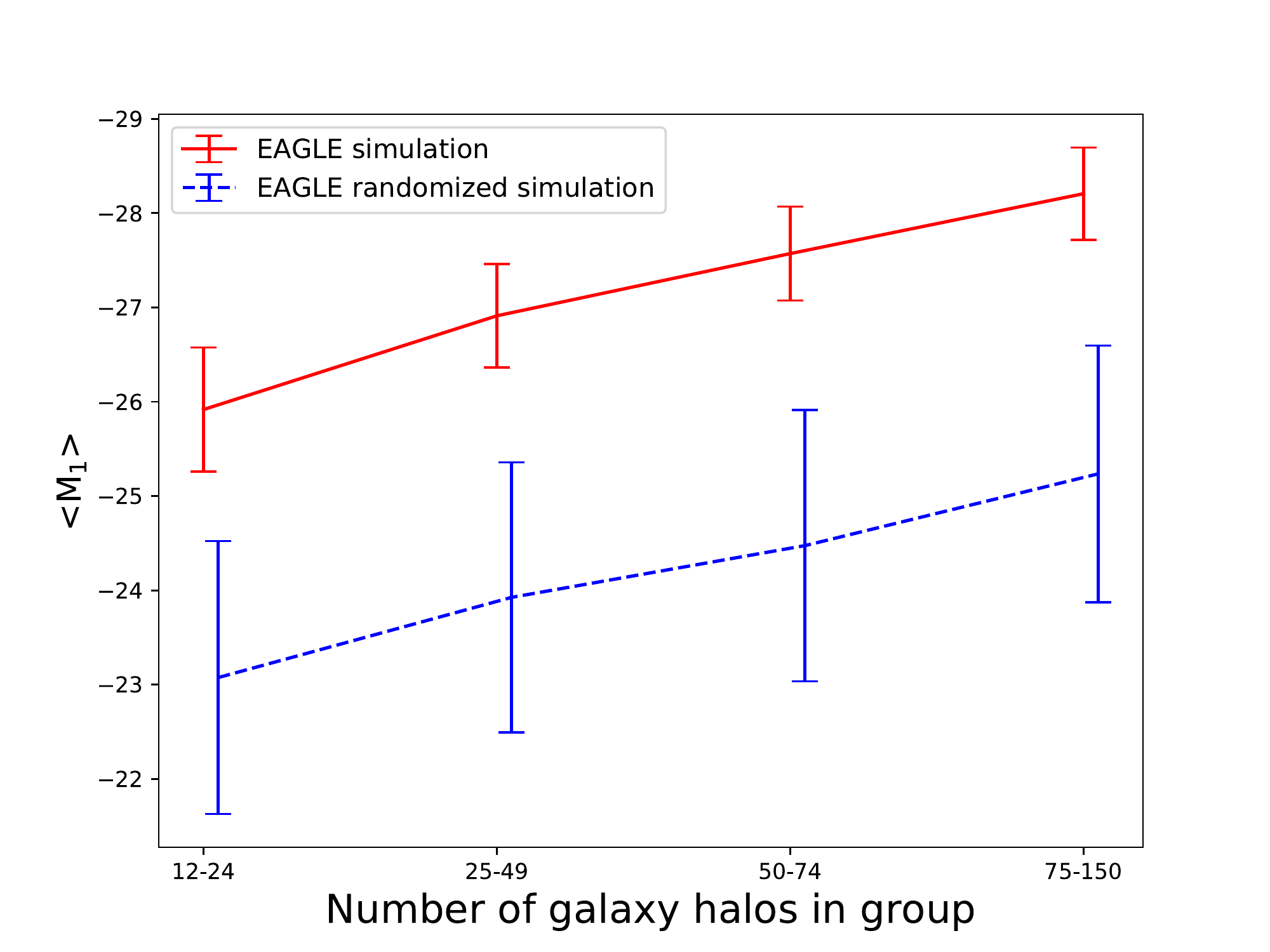}
 \caption{Average first-ranked magnitude proxy of the dark matter mass as a function of number of galaxy members in EAGLE dark matter only simulation.
 \label{fig:dm2} }
\end{figure}

We show in Figures~\ref{fig:dm1} and \ref{fig:dm2} and in Tables~\ref{tab:dm1} and \ref{tab:dm2} 
the results from the EAGLE dark matter only simulations.
The results show that the typical gap in
mass (expressed in magnitudes) even larger than in the far more complicated full baryonic
simulations and the values of $t_1$ even smaller ($t_1 = 0.16 \pm 0.02$). As noted
earlier a piece of this effect is due to definitions, not physics: matter tidally torn
off subhalos is, by definition, added to the parent halos, increasing the gaps. But it is
unlikely that this accounts for the whole effect. Therefore whatever
physical processes produce the large gaps seem to be even stronger in purely gravitational
simulations. We also show the randomized dark matter simulations as red lines in
Figure~\ref{fig:dm1} and \ref{fig:dm2}, giving the value of $t_1$ in Table~\ref{tab:dm2}.
These completely match the  \cite{1977ApJ...212..311T} statistical expectations.
The gap is larger in the original data than in the randomized data in these dark matter
simulations by $2.11 \pm 1.91$ magnitudes, even larger than in the galaxy simulations.

The results from IllustrisTNG dark matter only simulations are summarized 
in Tables~\ref{tab:dm3} and \ref{tab:dm4}. Again, we have a large gap between first and
second massive systems, and the gap is much larger in the original data than 
in the randomized data.

These results provide dramatic evidence that whatever causes the large gap (``too big to fail'') in observed
phenomenon is gravitational/dynamical in origin, since it is stronger in the dark matter only simulations
than in either the real or simulated works. Merging is a plausible explanation but more work would 
need to be done to prove this.

\begin{figure}
 \includegraphics[width=\columnwidth]{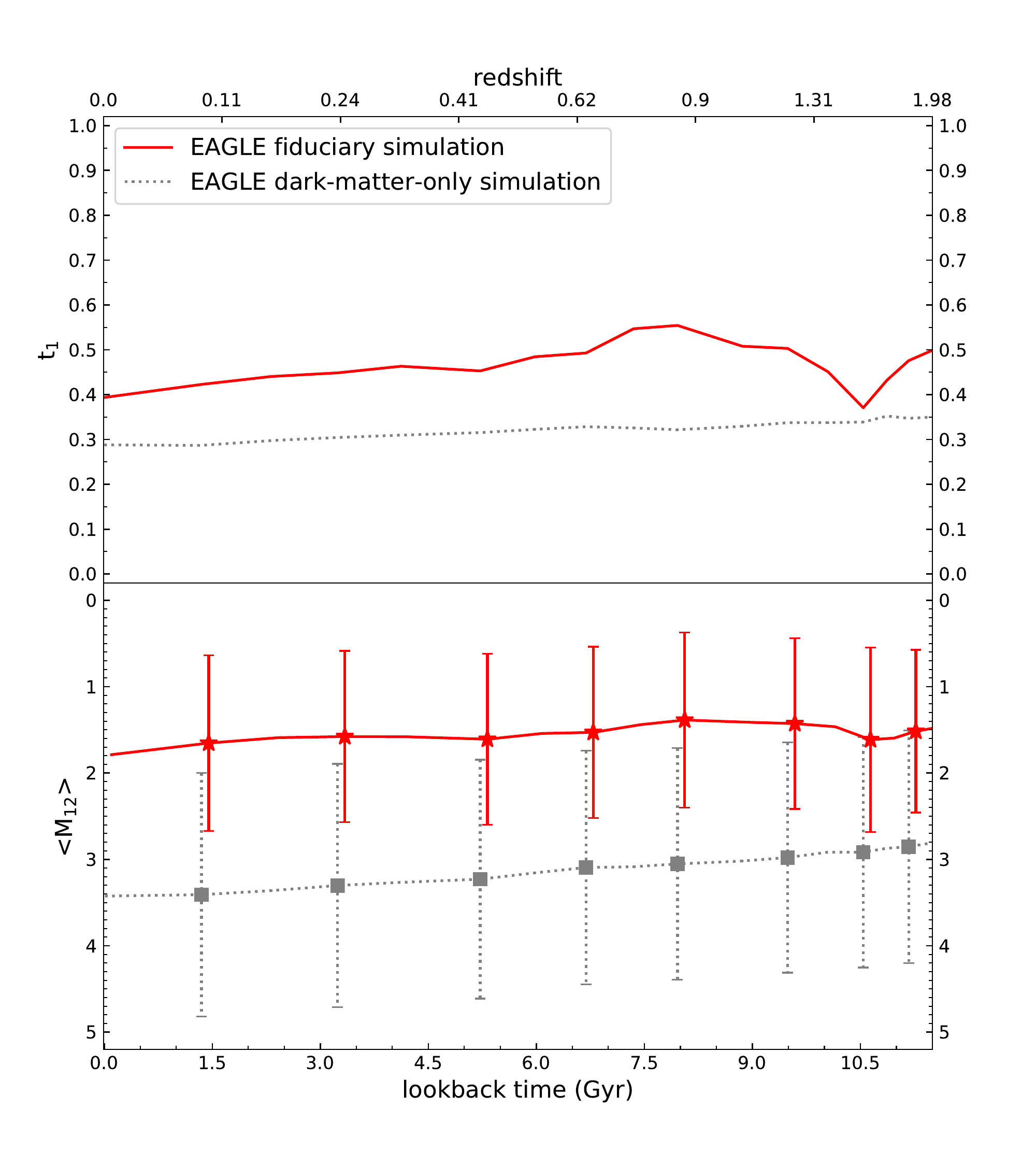}
 \caption{Evolution of $t_{1} $ parameter ($t_{1} = {\sigma(M_{1})}/{\langle  M_{12}\rangle}$) and the magnitude gap $\langle M_{12} \rangle$ in EAGLE simulation from $z=2$ to 0.
 \label{fig:evol} }
\end{figure}

 \section{Conclusion}\label{sec:summary}
We summarize the real data and the $\Lambda$CDM simulations in Table~\ref{tab:summary}.
We see that the $\langle M_{12} \rangle$ gaps are actually larger in the $\Lambda$CDM
simulations than in observed data and the anomalous $t_1$ statistic is as low in the 
simulated data as in the real data. The dark matter only simulations are even more
extreme.
 
First it is clear that modern simulations by active groups do not have a ``too big to fail''
problem. The gaps $\delm12$ in their simulated groups are large and the $t_1$ statistics
derived from their simulations can be even lower than those seen in real observational 
data. This is good news. Standard CDM simulations do not have a too big to fail problem.

Second, when their data in randomized \--- keeping the luminosity function constant \---
the gaps disappear and the data satisfies the statistical expectations with $t_1 > 1$. Therefore,
solving the problem was not based on particular feedback schemes which alter the luminosity
function, but rather it must be due to physical interactions in the groups and clusters. We noted
several physical interactions which would tend to produce the observed gaps and the
additional experiments that we did help to pick the winner.

In addition satellites moving through the gas in groups and clusters can be seen to be
losing material by ram pressure stripping and this effect, which we cannot easily quantify,
must also lead to an increase in the gap $\langle \delm12 \rangle$ and a lowering of $t_1$.

But one experiment that we performed in Section~\ref{sec:dmonly} showed us the dominant
physical mechanisms. We looked at dark matter only simulations from EAGLE and 
IllustrisTNG and found the gap (expressed in magnitudes) to be  $\langle M_{12} \rangle  
= 3.43 \pm 1.44$ and $\langle M_{12} \rangle  = 3.33 \pm 1.35$ much larger than in 
randomized dark matter systems and the $t_1$ statistic was $t_1 = 0.17 \pm 0.02$ and 
$t_1 = 0.19 \pm 0.04$, even lower than in the baryonic simulations
or the real data.

Since the sole physics acting in the dark matter experiments was based on gravity
and dynamics, it is clear that none of the complicated ``baryonic'' effects \--- including the first
two mentioned in this section \--- can be dominant in causing the large gaps and low value
of $t_1$.

Dynamical friction and the induced ``cannibalism'' can certainly produce the effects seen in 
the dark matter simulations so it is tempting to consider ``mergers'' to be the driving force
in groups and clusters leading to the big gaps and small values of $t_1$ seen in both the
baryonic and dark matter only simulations. A primitive numerical test of this was performed 
by \cite{1977ApJ...217L.125O} with promising results, but there is a strong argument on
the other side.

The total halo mass in solar type stars in our Milky Way (MW) is estimated by GAIA (and others)
observations not to exceed 1-2 percent of the mass in the MW disk 
\citep{2018A&A...616A..11G}. Since a large fraction
of the stars in any merging system would ultimately be found in the halo of a disk galaxy, that
tells us that whatever systems merged with our galaxy must not have weighed, more than a few 
percent of MW system. Here of course we are only considering the stellar component. 
This is true for other edge-on similar observed galaxies such as
NGC 4565 and for simulations as well. Using Illustris simulations \cite{2016MNRAS.458.2371R}
estimate the fraction of ex-situ stars and they find roughly 10 percent for MW size systems in 
other published work with perhaps half that much in their own simulations.

Thus both observations and simulations indicate that major mergers of stellar systems of
MW scale are rare. Hence explaining the group properties of these systems in terms of
merging stellar systems seems misguided. This argument applies to systems with total
halo mass less than $10^{12.5} \Msun$. There are multiple lines of evidence, however
in systems with total mass larger than $10^{13} \Msun$ that mergers can be significant.
The possibility remains, however, that mergers {\it before} significant star formation has 
occurred could be important and could explain the well established gaps in the luminosity 
functions seen in normal groups and clusters \--- both in the real and simulated worlds. 
Further work must be done to test this possibility.

But what is clear from the analysis presented in this paper is that ``too big to fail'' is not
a problem in the CDM scenario (nor, in all probability in the variant competitors) because
normal gravitational interactions within groups increase the mass of the most massive galaxy,
decrease the mass of the second ranked system and tend to produce large gaps.

\begin{acknowledgements}
We thank the anonymous referee for very helpful comments on the manuscript. 
We thank Gohar Dashyan, Daniel DeFelippis and Scott Tremaine for helpful discussions. 
Numerical simulations were run on the computer clusters of the Princeton Institute 
of Computational Science and engineering.
 \end{acknowledgements}

\bibliography{references}

\end{document}